\RequirePackage{fix-cm}
\documentclass[smallextended]{svjour3}       
\smartqed  
\usepackage{appendix}
\usepackage{graphicx}
\usepackage{fancyhdr}
\usepackage[final]{changes}
\usepackage{amsmath}
\usepackage{rotating}
\begin{document}
	
	\title{Assessing Systemic Risk in the Insurance Sector via Network Theory}
	
	
\author{Gian Paolo Clemente         \and
	Alessandra Cornaro
}


\institute{Gian Paolo Clemente \at
	Dipartimento di Matematica per le Scienze economiche, finanziarie e attuariali \\
	Università Cattolica del Sacro Cuore, Milano, Italy\\
	\email{gianpaolo.clemente@unicatt.it}         
	\and
	Alessandra Cornaro \at
	Dipartimento di Matematica per le Scienze economiche, finanziarie e attuariali \\
	Università Cattolica del Sacro Cuore, Milano, Italy\\
	\email{alessandra.cornaro@unicatt.it}}
	\date{Received: date / Accepted: date}
	
	\fancyhead[RO,LE]{\small\thepage}
	\fancyhead[LO]{\small Assessing systemic risk in the Insurance Sector via Network Theory}
	\fancyfoot[L,R,C]{}
	\maketitle
	
	\begin{abstract}
		We provide a framework for detecting relevant insurance companies in a systemic risk perspective. Among the alternative methodologies for measuring systemic risk, we propose a complex network approach where insurers are linked to form a global interconnected system. We model the reciprocal influence between insurers calibrating edge weights on the basis of specific risk measures. Therefore, we provide a suitable network indicator, the Weighted Effective Resistance Centrality, able to catch which is the effect of a specific vertex on the network robustness. By means of this indicator, we assess the prominence of a company in spreading and receiving risk from the others. 
				
		\keywords{Systemic risk \and Insurance market \and Robustness and vulnerability of complex network \and Market-based risk measures}
		\noindent\textbf{JEL classification code}{: G01, D85, G32, G28.}
		
	\end{abstract}
	
	\section{Introduction}
	\label{intro}
	
	The concept of systemic risk has gained increasing attention from both regulators and academics. Since 2009, a regulatory response to the revealed vulnerability of the banking sector in the financial crisis of 2007–08, and an attempting to find a solution to solve the \lq\lq too big to fail\rq\rq interdependence between global systemically important banks (G-SIBs) and the economy of sovereign states have been observed. In particular, the Financial Stability Board (FSB) started to develop a method to identify G-SIBs to whom a set of stricter requirements would apply (see \cite{FSB2010}, \cite{FSB2011} and \cite{FSB2013}). A first official version of the G-SIB list was published by FSB in November 2011, and it has ever since been updated each year. \\
	The debate regarding the systemic nature of financial institutions has been ongoing for many years and also involved the insurance sector. The  International  Association  of  Insurance  Supervisors  (IAIS) played an  important  role  in  this  global  initiative. The IAIS (\cite{IAIS} and \cite{IAIS2011}) suggested that the specificities of the insurance activity should be  duly considered  when  attempting  to  extend  the definition of systemic risk to  the insurance  sector,  namely regarding  the  specificities  of  underwriting. Although it is recognized that insurance is a financial sector with  significant  links  to  the  real  economy,  it  differs  from  the  other  financial services by its business model (e.g. for the \lq\lq inverted cycle of production\rq\rq). However, under  the  leadership  and  steering  of the FSB, the IAIS has focused on the analysis of the potential systemic relevance of  insurers. 
	Both IAIS and several papers in the literature (see \cite{Acha}, \cite{Cummins}, \cite{Eling}, \cite{Geneva}, \cite{IAA}) concluded that, although   insurance   companies   are   less   prone   to   systemic   risk   and   less vulnerable  than  banks,  some  non-traditional  activities  may  entail  some  risk, namely due to the high leverage and implied guarantees associated with them.
	For  this  purpose,  the  IAIS  has  developed  a  methodology  to  assess and ultimately identify global systemically important insurers (G-SIIs), as well as a range of policy measures to be applied to them (\cite{IAIS2013}, \cite{IAIS2013b}). An initial list of nine multinational insurance groups that have been classified as G-SIIs has been released in 2013. These insurers were identified on the belief that should one of them become insolvent and fail in a disorderly manner, it could have negative impacts on the stability of the global financial system. The IAIS assessment and FSB identification became an annual process, with the FSB and IAIS developing a framework of policy measures to be applied to G-SIIs with the objective of reducing the negative externalities stemming from the potential disorderly failure posed by a G-SII. This framework of G-SII policy measures consisted of three main elements: higher loss absorbency requirements; enhanced group-wide supervision and group-wide recovery, resolution planning and regular resolvability assessments. \\
	In November 2019, the IAIS (\cite{IAIS2019}) finalized a holistic framework for the assessment and mitigation of systemic risk in the insurance sector. The framework moves away from taking solely an entities-based approach and sets out an activities-based approach for sector-wide risk monitoring and management, as a key component of the framework, and tools for dealing with the build-up of risk within individual insurers or groups of insurers. The goal is that an appropriately implemented framework will provide an enhanced basis for mitigating systemic risk in the insurance sector. In this context, the annual identification of G-SII has been suspended and, in November 2022, the FSB will review the need to either discontinue or re-establish an annual identification of G-SIIs on the basis of the initial years of implementation of the holistic framework. \\
	One of the key points highlighted by this document is the prominent role of interconnections in the financial system and the importance of measuring them in terms of risk. Also within the extensive and variegated literature on systemic risk measurement in the banking sector,  modeling the interconnections plays a leading role. On the one hand,  statistical market-based measures of individual risk such as value at risk (VaR) and expected shortfall (ES), which are traditionally employed in risk management but are also used for regulation, have evolved to account for tail-dependence among banks (see, e.g., \cite{Benoit2017}, \cite{Bisias2012} and \cite{DeBandt2013}) for recent surveys in this field). On the other hand, the structure of financial and economic systems, which are characterized by interacting agents, can be described as a complex system and network indicators have been used to assess systemic risk in the banking sector (for recent surveys about networks and systemic risk, we refer to \cite{Caccioli2018} or \cite{Neveu2018}). Additionally, several works focused on the use of network indicators for identifying the most relevant banks in the system (see, e.g., \cite{Clemente2020a}, \cite{RoviraKaltwasser2019}, \cite{SpeltaPEcora}). \\
	Less attention has been paid in the literature to the measurement of systemic risk in the insurance sector and to the identification of most relevant insurers. A model for systemic risk of insurance companies and banks has been provided in \cite{Acha2016}. A rational for the macro-prudential regulation of insurance companies has been described in \cite{Gomez}. In \cite{Cummins2014} systemic risk for the U.S. insurance sector has been evaluated, focusing on the interconnectedness of the system. Systemic relevance of the European insurance industry has been instead explored in \cite{Bertin}. The importance related to the identification of G-SIIs has been stressed in \cite{Guine}. In \cite{Denko} copula-garch models combined with minimum spanning trees have been proposed and then the contribution of each institution has been analyzed using CoVaR  (see \cite{Adrian2016}). In \cite{Fung} it has been showed that the recent policy measures proposed by the IAIS are perceived by the market as heading in the right
	direction and concluded that the regulation partially achieved its intended objective of reducing the systemic risk of G‐SIIs. A review of different approaches is given in \cite{Job}.\\
	In this context, we propose to identify the most relevant insurers in the market, combining the network approach with the standard risk measures that are based on market data. In this way, we integrate the full picture that is provided by network analysis and the forward-looking approach of market-based statistical measures. 
	We model the insurance market via a weighted network in which vertices represent firms and edges are quantified via market-based measures. Each edge is indeed weighted considering the expected shortfall of the company and the marginal expected shortfall (MES), proposed by \cite{Acharya2017}, computed on each pair of firms. In this way, the weights allow to exploit the tail relations between couple of insurers.
	To identify most relevant companies in the market, we make use of a suitable network indicator, namely the Effective Resistance Centrality, defined in \cite{Clemente2020} for unweighted networks as the relative drop of the Effective Graph Resistance (also known as Kirchhoff index) caused by the removal of a vertex from the network. In particular, we extend here the version provided in \cite{Clemente2020} to the weighted case. The proposed indicator appears suitable for identifying critical vertices in the network and, therefore, relevant firms in the market. Indeed, the Kirchhoff index has been widely used for assessing the robustness of the network. For instance, it has been showed in \cite{Ellens} and \cite{WangP} that this indicator allows to assess the ability of a network to continue performing well when it is subject to failure and/or attack. Therefore, vertices whose removal mainly affects the value of the Kirchhoff index, are critical in terms of vulnerability of the network. By means of this analysis, we focus on two aspects of the contagion, namely, the local level of interconnection and the dominant position of a firm in the system in terms of spreading or receiving risk. \\
	The results show that the networks well capture the behavior of interconnections over time. The pattern of networks' weights appear to be strictly related to the financial condition of the market. Furthermore, current G-SIIs are identified by this network indicator with a few exceptions. Indeed, this type of approach can be considered a complement to the more traditional approaches that are based on balance sheet and regulatory data.\\
	The remainder of the paper is organized as follows. In Section \ref{sec:Pre} we provide some preliminaries about graph theory and weighted Kirchhoff index. In Section \ref{sec:met}, we explain how the network and the weights of edges are obtained. We also define the Weighted Effective Resistance Centrality that can be interpreted as an extension to weighted networks of the indicator provided in \cite{Clemente2020}. In Section  \ref{sec:res}, we present and discuss the results that were obtained for the sample of insurance companies over the period from 2001 to end of 2015. Section \ref{sec:conc} concludes.

\section{Preliminaries}
\label{sec:Pre}
\subsection{Graph theory}
Let us firstly recall some standard definition and results about graph theory (see, for more details, \cite{Bollobas1998}, \cite{Grone1994} and \cite{Harary}).

\noindent A graph $G=(V,E)$ is a pair of sets $(V,E)$, where $
V=\left\{ 1,...,n\right\} $ is the set of vertices and $E\subseteq
V\times V$ is the set of edges. We consider graphs with fixed order $\left\vert
V\right\vert =n\ $ and fixed size $\left\vert E\right\vert =m.$\\
An undirected graph is a graph in which if $(i,j) \in E$ , then $(j,i) \in E$, whereas a directed graph (digraph) is a graph in which each edge (arc) is an ordered pair of vertices. We denote with $e_{i,j}$ the edge connecting vertices $i$ and $j$. When two vertices share an edge, they are called adjacent. A loop is a degenerate edge of a graph which joins a vertex to itself (also called a self-loop). Multiple edges are two or more edges connecting the same two vertices $i$, $j$. In the next we will neglect self-loops and multiple edges. 
Moreover, a weight $w_{ij}$ is possibly associated to each edge $e_{i,j}$, in this case we will have a weighted (or valued) graph.

A simple graph is an unweighted, undirected graph containing no graph loops or multiple edges. 
\noindent	A $i-j$ path is a sequence
of distinct adjacent vertices from vertex $i$ to vertex $j$. The distance $d(i,j)$ between $i$ and $j$ is the length of the shortest path joining them when such a path exists, and it is set to $+\infty$ otherwise. A graph $G$ is connected if there is a path between every couple of vertices.
Let $\mathbf{\pi}
=(d_{1},d_{2},..,d_{n})$ denote the degree sequence of $G$ arranged in non-increasing order $d_{1}\geq d_{2}\geq \cdots \geq d_{n}$, being $d_{i}$ the
degree of vertex $i$.

A non-negative $n$-square matrix $\mathbf{A}$, representing the adjacency relationships between vertices of $G$, is associated to the graph (the adjacency matrix); the off-diagonal elements $a_{ij}$ of $\mathbf{A}$ are equal to 1 if vertices $i$ and $j$ are adjacent, 0 otherwise;
if the graph has self-loops the corresponding diagonal elements of A are equal to 1. 
Given the diagonal matrix $\mathbf{D}$ of vertex degrees, the matrix $%
\mathbf{L}=\mathbf{D}-\mathbf{A}$ is known as the Laplacian matrix of $G$.
For a graph $G$ with non-negative edge weights $w_{ij}$  we denoted by $\mathbf{W} = [w_{ij}]$ the weighted adjacency matrix and the weighted Laplacian matrix by $\mathbf{L}^{W}=\mathbf{S}-\mathbf{W}$ where $\mathbf{S}$ is the diagonal matrix of strengths, with elements
$\mathbf{S}_{ii}=s_i$. To summarize, the elements of the weighted Laplacian are given by
\begin{equation}
	\mathbf{L}_{ij}^W=\left\{ 
	\begin{array}{ccc}
		s_{i}=\sum_j w_{ij}  & 
		\text{ if } & i=j \\ 
		-w_{ij}
		& \text{ if } & (i,j) \in E \\
		
		0  & 
		& \text{otherwise} 
	\end{array}.
	\right. 
\end{equation}%
The weighted Laplacian  is  a  positive  semidefinite  matrix  i.e.   the  eigenvalues  of  the  Laplacian  are  nonnegative, hence $\mu^{W} _{1}\geq
\mu^{W} _{2}\geq ...\geq \mu^{W} _{n}=0$. \\


\subsection{The weighted Kirchhoff index and the normalized weighted Kirchhoff index}
The \textit{Kirchhoff index} $K(G)$ of a simple connected graph $G$ was defined by Klein
and Randi\'{c} in \cite{Klein} as the accumulated effective resistance between all pairs of vertices.
In addition to its original definition, the Kirchhoff index can be rewritten as
$K(G)=n\sum_{i=1}^{n-1}{\frac{1}{{\mu_{i}}}}$, in terms of the eigenvalues of the Laplacian matrix $\mathbf{L}$ (see \cite{Gutman}, \cite{Zhu}).
For a weighted and undirected graph we have the weighted Kirchhoff Index defined as (see \cite{Klein}):
\begin{equation}
	K^{W}(G)=n\sum_{i=1}^{n-1}{\frac{1}{{\mu^{W}_{i}}}}.  \label{Wkirchhoff}
\end{equation}

We recall some important results given in \cite{Klein}:
\begin{theorem}
	
$K^{W}(G)$ is a non-increasing function of the edge weights, $K^{W}(G)$ does not increase when edges are added.	
\end{theorem}	
The following corollary follows from the last theorem:
\begin{corollary}
	The total effective resistance strictly decreases when edges are added or weights are increased.
\end{corollary}

In order to compare the value of the Kirchhoff index for networks with different orders, we can consider the \textit{normalized weighted Kirchhoff index}. Assuming $w_{i,j} \in [0,1]$, we can define it in the following way extending to the weighted case the indicator defined in \cite{Wang}:

\begin{equation}
	K^{W}_{N}(G)=\frac{K^{W}(G)}{\binom{n}{2}}, \label{Normkirchhoff}
\end{equation}%
where the denominator considers the maximum sum of the weights.

It is worth pointing out that the Kirchhoff index can be highly informative
as a robustness measure of a network, showing the ability of a network to continue performing well when it is subject to failure and/or attack. In fact, the pairwise effective
resistance measures the vulnerability of a connection between a pair of vertices that
considers both the number of paths between the vertices and their length. A small value
of the Kirchhoff index therefore indicates a robust network. 

\section{Modeling the insurance system} \label{sec:met}

In this section, we model the insurance market as a directed weighted network, which is
formally defined as a weighted graph with $n$ vertices, representing insurance companies in different countries, and $m$ edges, representing the relations between firms. A positive weight, denoted by $w_{ij}$, is associated with each edge $(i,j)$.  \\
In Section 3.1 we define how the weights on the edges are calibrated using market-based risk measures. In Section 3.2 we provide a new network indicator useful to assess the relevance of each firm in terms of robustness of the network.

\subsection{Risk network of the insurance market}
\label{sec:1}
Our purpose is to build a specific risk network that summarizes the tail dependences between insurance companies by considering specific risk measures. The \lq\lq tail impact\rq\rq between financial institutions is indeed used in the construction of a network for the insurance system by building on standard market-based measures of systemic risk. These measures are estimated on equity returns; therefore our model shares various features with correlation networks that are applied to the financial sector and their evolution (e.g. \cite{Billio2012} and \cite{Kenett2012}). \\
In particular,  we are interested in quantifying the impact of the distress of an insurance company on the others. This point of view is consistent with the definition of systemic risk. The proposed approach is to use mean expected shortfall (MES) as a measure of the tail influence of a company on another (as proposed in \cite{Clemente2020a}). In \cite{Acharya2017}, MES for a firm $i$ is defined in terms of the return, which is denoted as $r_i$, as its expected return conditional on the return of the system $r_{s}$ being below its Value-at-Risk (VaR) level:

\begin{equation}
MES_{i}=-E(r_{i}|r_{s}\leq VaR_{s}).
\end{equation}

It is noticeable that the definition is similar to the CoVaR, which instead quantifies the risk of the system conditional on the distress of a single firm. In this work, we consider instead pairs of insurers and we define the expected shortfall of an insurer $i$ conditional on insurer $j$ being in distress as:

\begin{equation}\label{MES}
MES_{i|j}=-E(r_{i}|r_{j}\leq VaR_{j}).
\end{equation}

In particular, we consider the difference $ES_{i}-MES_{i|j}=-E(r_{i}|r_{i}\leq VaR_{i})+E(r_{i}|r_{j}\leq
VaR_{j})$ to quantify the tail influence of an insurer $i$ on the insurer $j$. This difference is always non-negative and values that are closer to zero indicate a stronger tail impact of the insurer $j$ on insurer $i$. In other words, we interpret previous difference as the risk of the insurer $i$ that is not driven by insurer $j$. To assure comparability among institutions, we scale this quantity by a measure of the risk for the insurer $i$, i.e., the difference between the unconditional expected return and the tail expected return. We define our measure of the impact of $j$ on 
$i$ as:

\begin{equation}\label{I}
I_{i|j}=\frac{-E(r_{i}|r_{i}\leq VaR_{i})+E(r_{i}|r_{j}\leq VaR_{j})}{%
E(r_{i})-E(r_{i}|r_{i}\leq VaR_{i})}.
\end{equation}

$I_{i|j}$ is always non-negative since both the numerator and the denominator are non-negative. It is noteworthy that
in case the inequality $E(r_{i})<E(r_{i}|r_{j}\leq VaR_{j})$ is satisfied, we have a positive reaction of $i$ to the distress of the company $j$.  Our purpose is to analyze systemic risk, therefore we neglect the positive impact obtaining values of $I_{i|j}$ bounded between 0 and 1. \\
Hence, we construct a network in which each insurance company is a vertex and weights on the edge are defined as follows:

\begin{equation}\label{weights}
	w_{ji}=\left\{
	\begin{array}{ll}
		1-I_{i|j} & \mbox{if } i \neq j \mbox{ and } E(r_i) \geq E(r_i|r_j\leq-VaR_j)\\
		0 & \text{otherwise}
	\end{array}
	\right.
	.
\end{equation}
It is worth pointing out that in formula (\ref{weights}), weights $w_{ji}$ reflect the impact of the insurer $j$ on insurer $i$. Since we are considering $1-I_{i|j}$ higher value of weight $w_{ji}$ means a high impact between institutions or, in other words, it corresponds to a lower portion of the risk not being driven by the company $j$.  

\subsection{A suitable network indicator: Weighted Effective Resistance Centrality}
\label{sec:Newm}
To identify the most significant insurance companies in the system, we propose a suitable local network indicator that is an extension to weighted networks of the vertex-based Effective Resistance Centrality provided in \cite{CCJOL} and \cite{Clemente2020}.
We now provide a definition and a structural description of the vertex-based Weighted Effective Resistance Centrality.

\subsection*{Vertex-based Weighted Effective Resistance Centrality}
Let $G=(V,E)$ be a weighted and connected graph of $n$ vertices and $m$ edges and $G_{i}$ the graph obtained by removing the vertex $i$ and all its related connections from $G$. The Weighted Effective Resistance Centrality  $R^{W}_{K}(i, G)$ of the vertex $i$ is defined as
\begin{equation} 
	R^{W}_{K}(i, G)=\frac{(\Delta K^{W}_{N})_{i}}{K^{W}_{N}(G)}=\frac{K^{W}_{N}(G_{i})-K^{W}_{N}(G)}{K^{W}_{N}(G)}.
	\label{R_v}
\end{equation}

In (\ref{R_v}) we consider at the numerator the drop of the weighted normalized Kirchhoff index in order to provide a consistent comparison between graphs $G$ and $G_{i}$ that have different orders.	

\noindent Notice that the quantity $(\Delta K^{W}_{N})_{i}$ is not always positive, depending on the relevance of the specific vertex $i$ in the network. This measure can be useful to identify strategic vertices, whose failure can affect the resilience of the network. Furthermore, the measure also allows to detect eventual vertices to be removed in order to improve the robustness of the network.

\section{Results and discussion}
\label{sec:res}
\subsection{Dataset Description and Preliminary Results}

In this section, we study the insurance market via a network approach. We selected largest insurance companies in terms of asset size and market capitalization.
The companies that belong to the sample are listed in Table \ref{ListBank} in Appendix. For each firm, equity returns have been collected\footnote{\added{The equity returns have been collected from Bloomberg.}} on a daily basis in the period that ranges from the beginning of January 2001 to the end of December 2015. \\
Returns  have  been  split  by  using  monthly  windows and have been used to construct a time-varying weighted network in each period. To compute weights as defined in (\ref{weights}), ES and MES at the 95\% confidence level are estimated for each firm via historical simulation. Additionally, the network has been symmetrized considering as edge weight the average effect between the pair of firms involved.

Hence,  for  each  window,  we  have  a  network  $G_{t}=  (V_{t},E_{t})$  (with $t=  1,...,180$),  where insurance companies are vertices and the weights of the edges are related to the impact between institutions. Notice that the number of assets can vary over time.  Indeed we have considered the 119 firms reported in Table \ref{ListBank}, but some of these firms have no information available for some specific time periods.  Therefore, in each window, we have considered only insurers whose observations are sufficiently large to ensure a significant estimation of weights. 

\begin{figure}[!h]
	\centering
	\includegraphics[width=5cm,height=5cm]{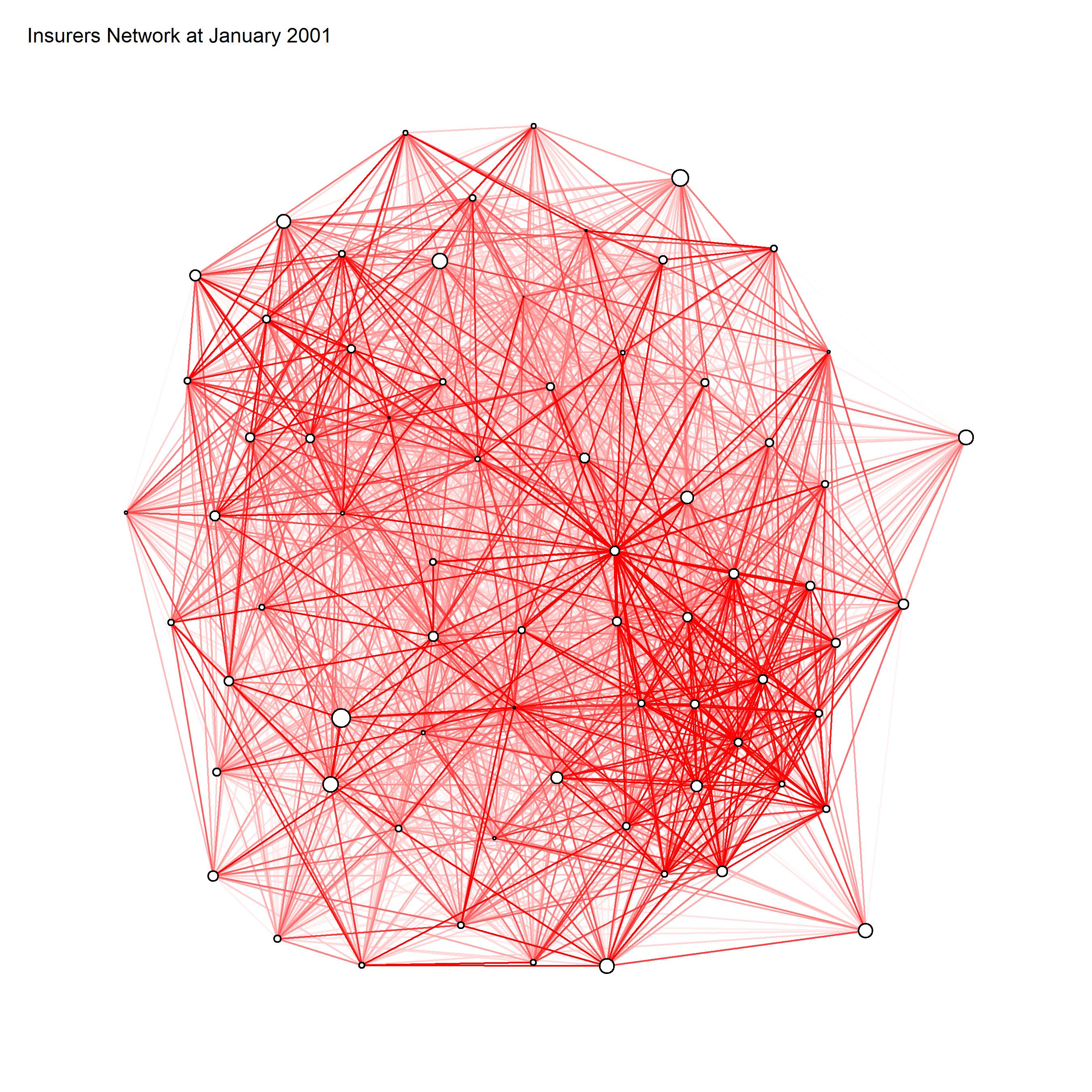}
	\qquad
	\includegraphics[width=5cm,height=5cm]{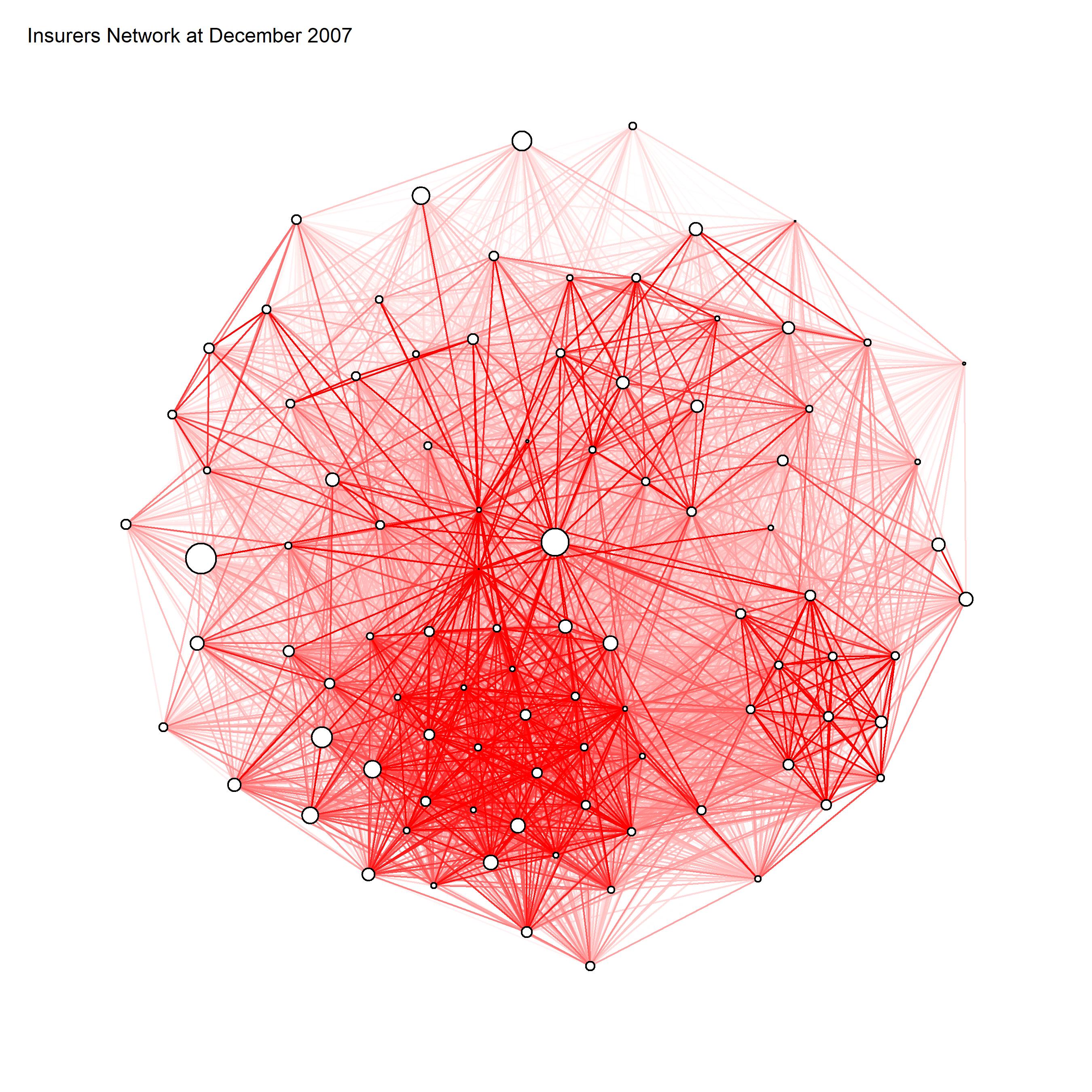}
	\includegraphics[width=5cm,height=5cm]{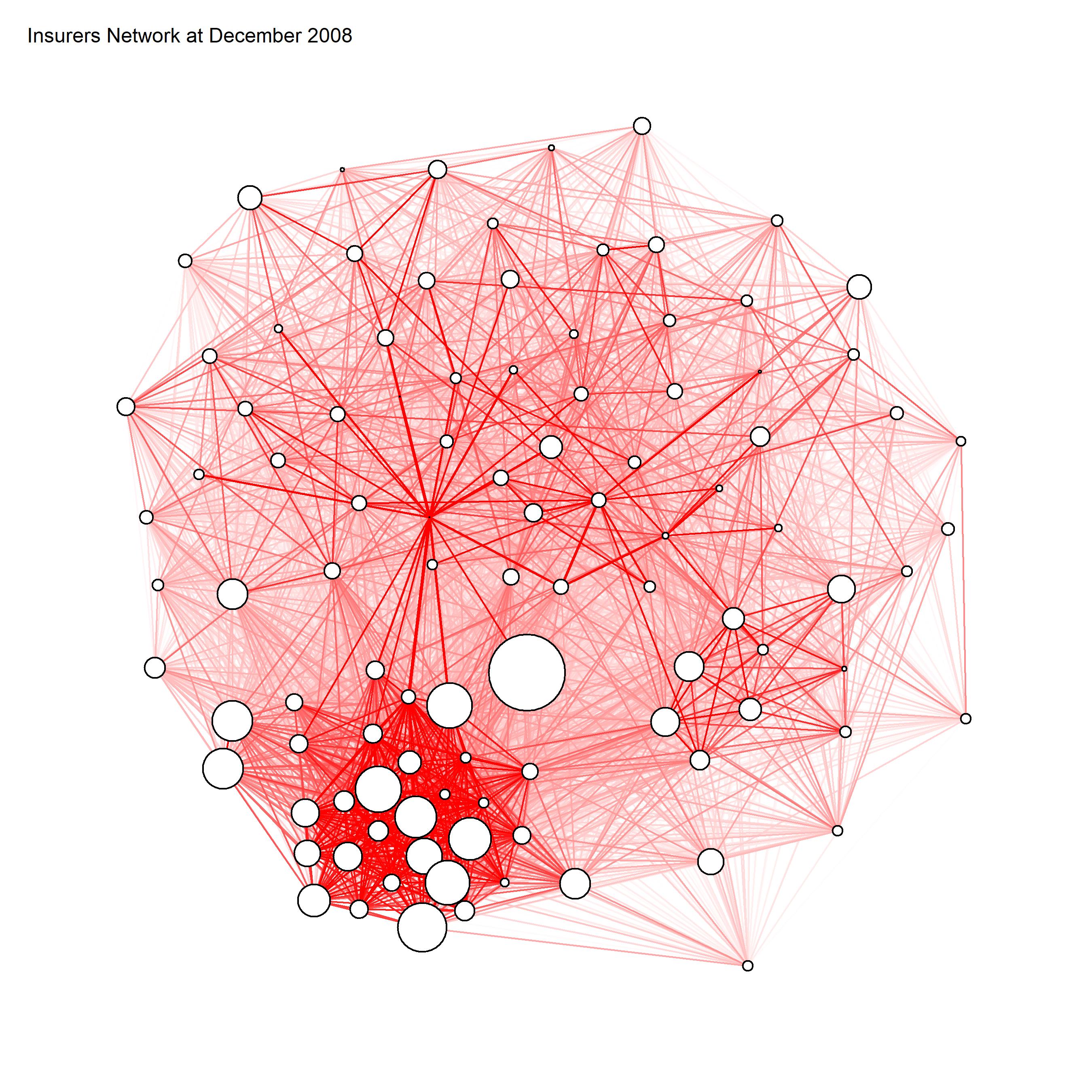}
	\qquad
	\includegraphics[width=5cm,height=5cm]{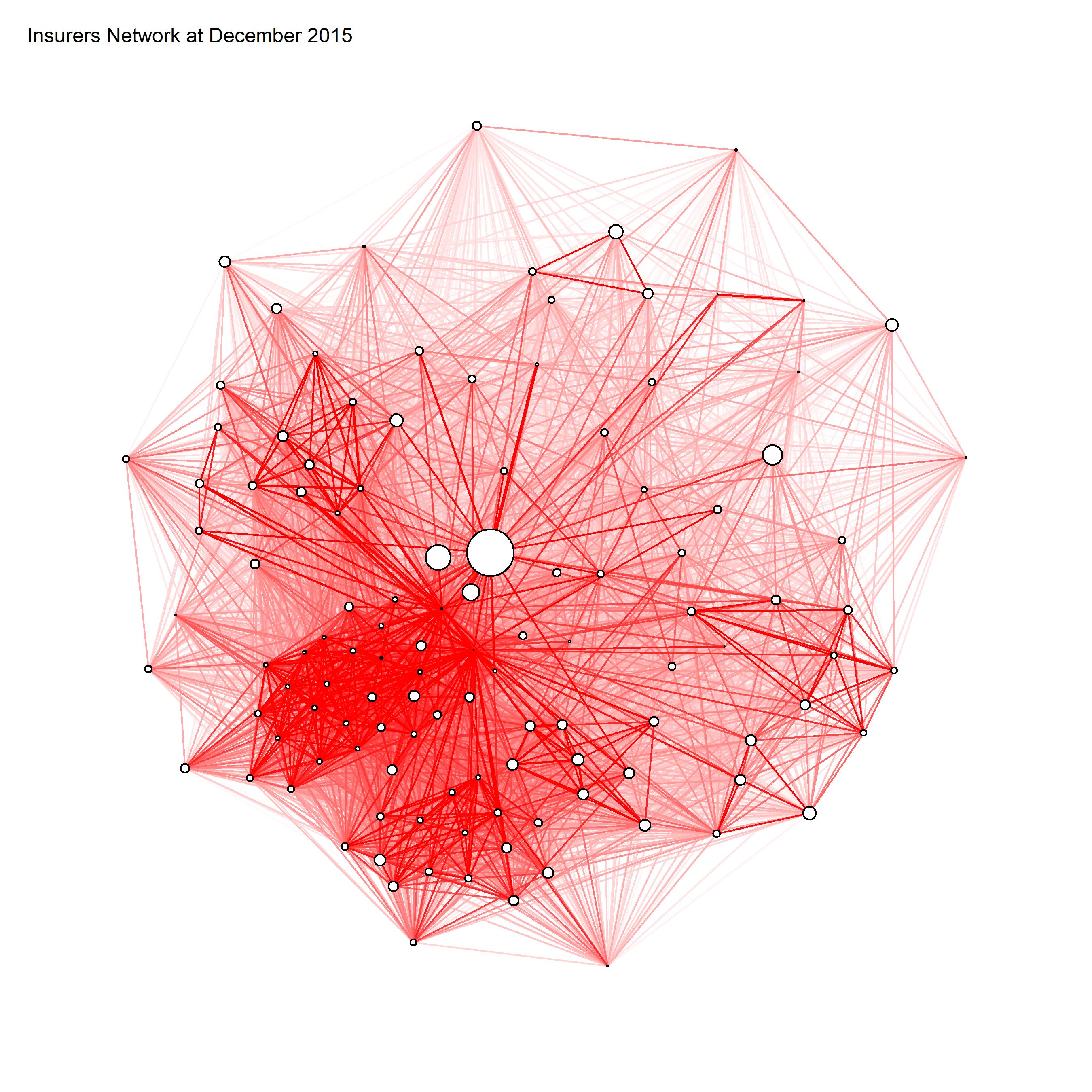}
	\caption{Networks of the insurance market in four time periods. The edge opacities are proportional to the weights (each of which represents the intensity of the impact of a firm on another firm). Size of vertices is proportional to the expected shortfall of the firm.}
	\label{fig:Networks}
\end{figure}

\begin{figure}[!h] 
	\centering
	\includegraphics[width=5.5cm,height=5.5cm]{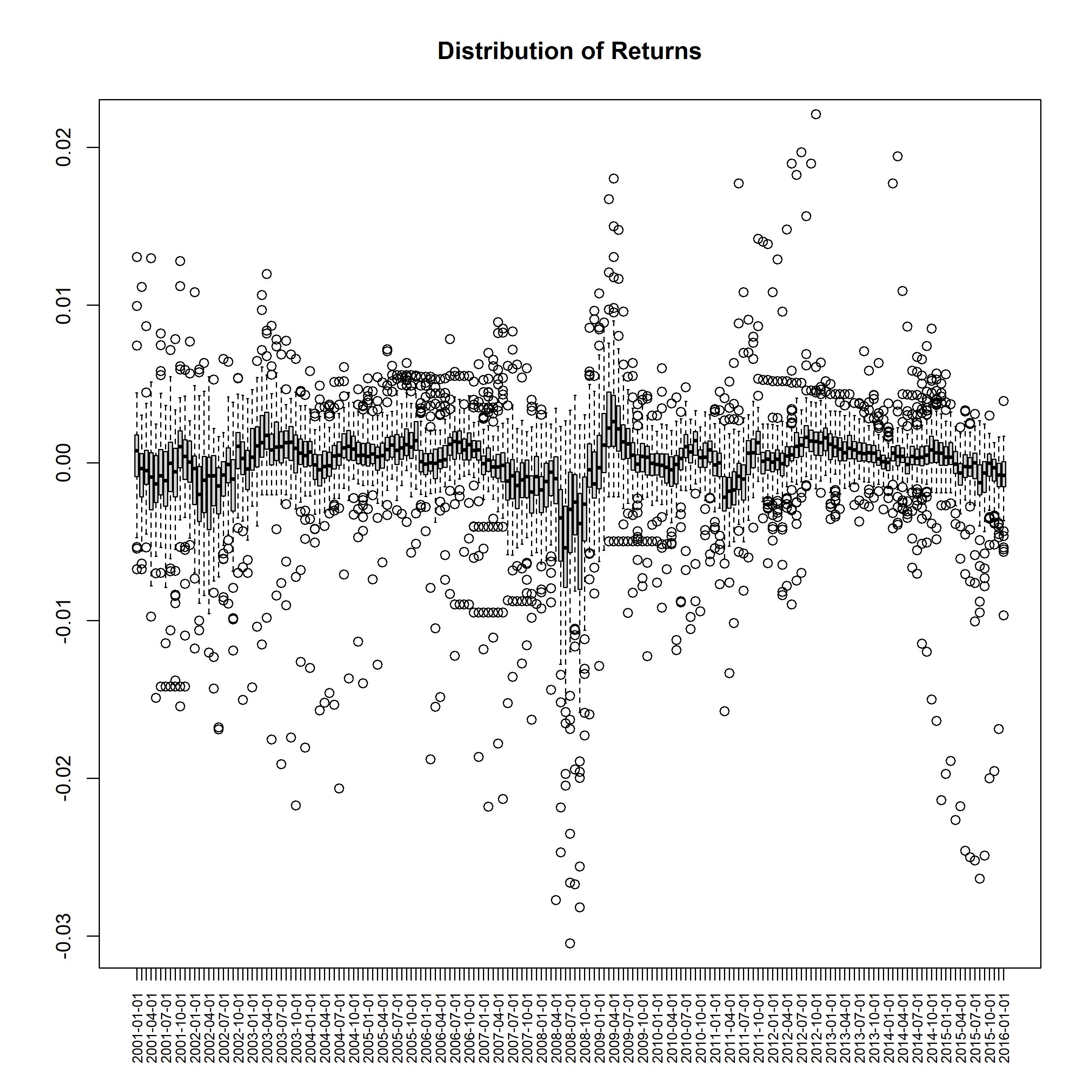}
	\qquad
	\includegraphics[width=5.5cm,height=5.5cm]{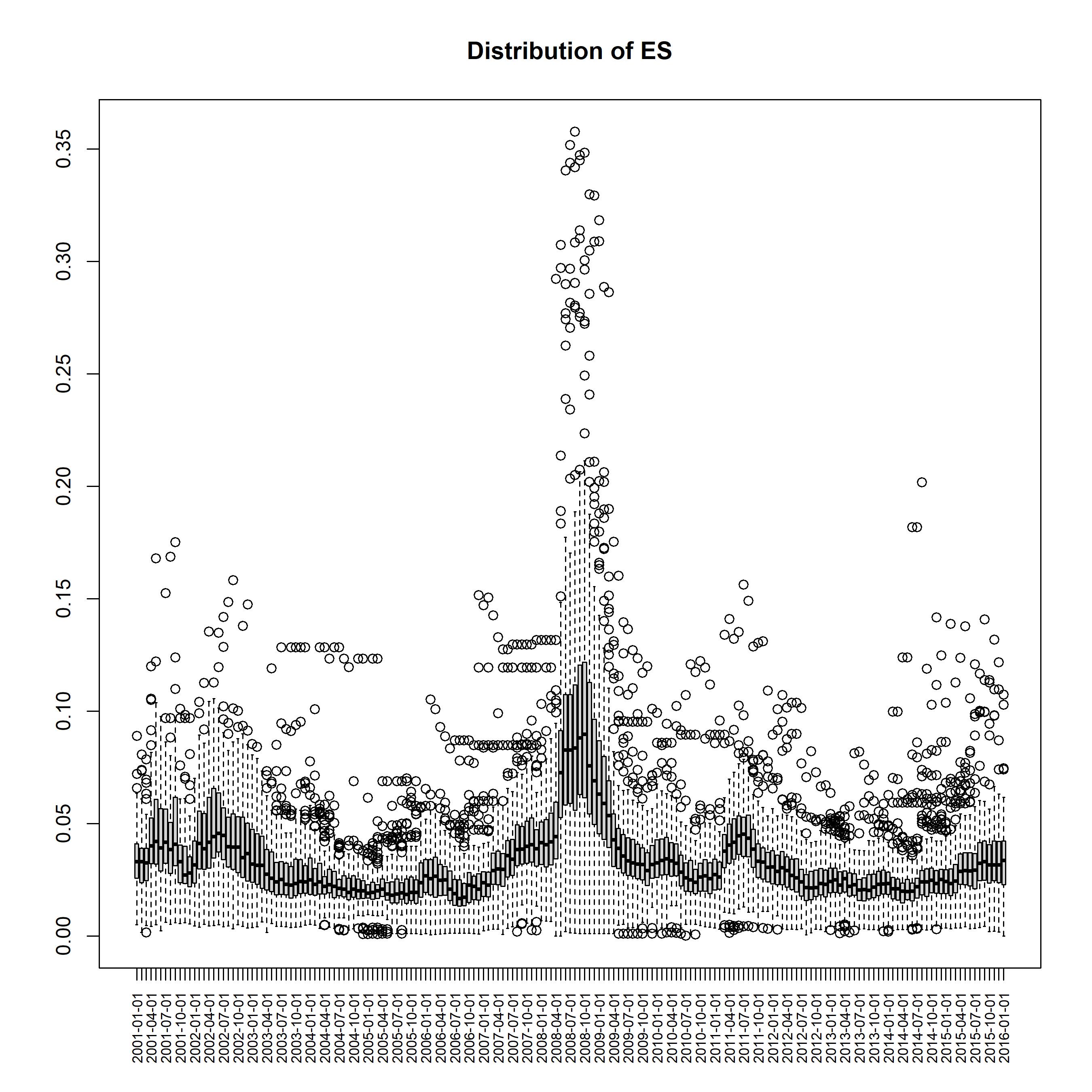}
	\caption{Distributions of average returns and expected shortfall at various time periods. Each distribution is displayed as a classical box-and-whisker plot. Each rectangle includes data between the first and third quartiles, while the horizontal line inside indicates the median value. The whiskers go from the first and third quartiles to 1.5 times the interquartile range (we are including data that are within 3/2 times the interquartile range). Observations that fall outside 1.5 times the interquartile range are plotted as dots.}
	\label{fig:Distr}
\end{figure}

In Figure \ref{fig:Networks}, we report four of the 180 networks that have been analyzed. As previously discussed, each vertex represents an insurance company and the weighted edge $e_{i,j}$ measures the average MES-based impact between a company $i$ and a company $j$. According to a preliminar visual inspection, the networks appear to be very dense. Indeed the density, which is computed as the ratio of the number of observed edges to the number of potential edges, is equal to 0.7 in the first network and to 0.81 at the end of the period (December 2015). Although the networks are densely connected, they are not complete because we set a weight equal to zero when minus MES exceeds the expected returns of insurance company $i$ in the same time period (see formula (\ref{I})).\\
Moreover, the networks are denser in periods of crisis, e.g., the networks covering data for years 2007 and 2008\footnote {Density exceeds 0.8 in these years with a maximum value of 0.88}, because of both lower returns and higher dependency (and, hence, higher MES) between companies\footnote{This point will become more evident later via analysis of Figure \ref{fig:Distr}}. In particular, it is noticeable the situation at the end of 2008 where a very dense subgraph is observed characterized by a subset of insurers (mainly U.S. firms) highly correlated and with high values of expected shortfall. \\

Main measures that are involved in the computation of $I_{i|j}$ are summarized in Figure \ref{fig:Distr}. We analyze how the distributions of the average returns and Expected Shortfall vary over time. \\
In this regard, the left figure shows the fluctuations of average returns of each company in the sample. As expected, in periods that are characterized by either the Lehman Brothers failure or by the sovereign debt crisis, the average returns are negative and there is higher volatility among firms. 
According to the ES distribution (Figure \ref{fig:Distr}, right), the 2007-2009 crisis is outstanding in terms of the size and frequency of the extreme daily market losses. During 2008 the median of the distribution reaches values close to 10.5\%, significantly higher than the values around 3\% observed in quiet periods. \\
Also the sovereign debt crisis period (in 2011) is characterized by a slight increase in the median of ES distribution (equal to approximately 5\%). \\

\begin{figure}[!h] 
	\centering
	\includegraphics[width=10cm,height=10cm]{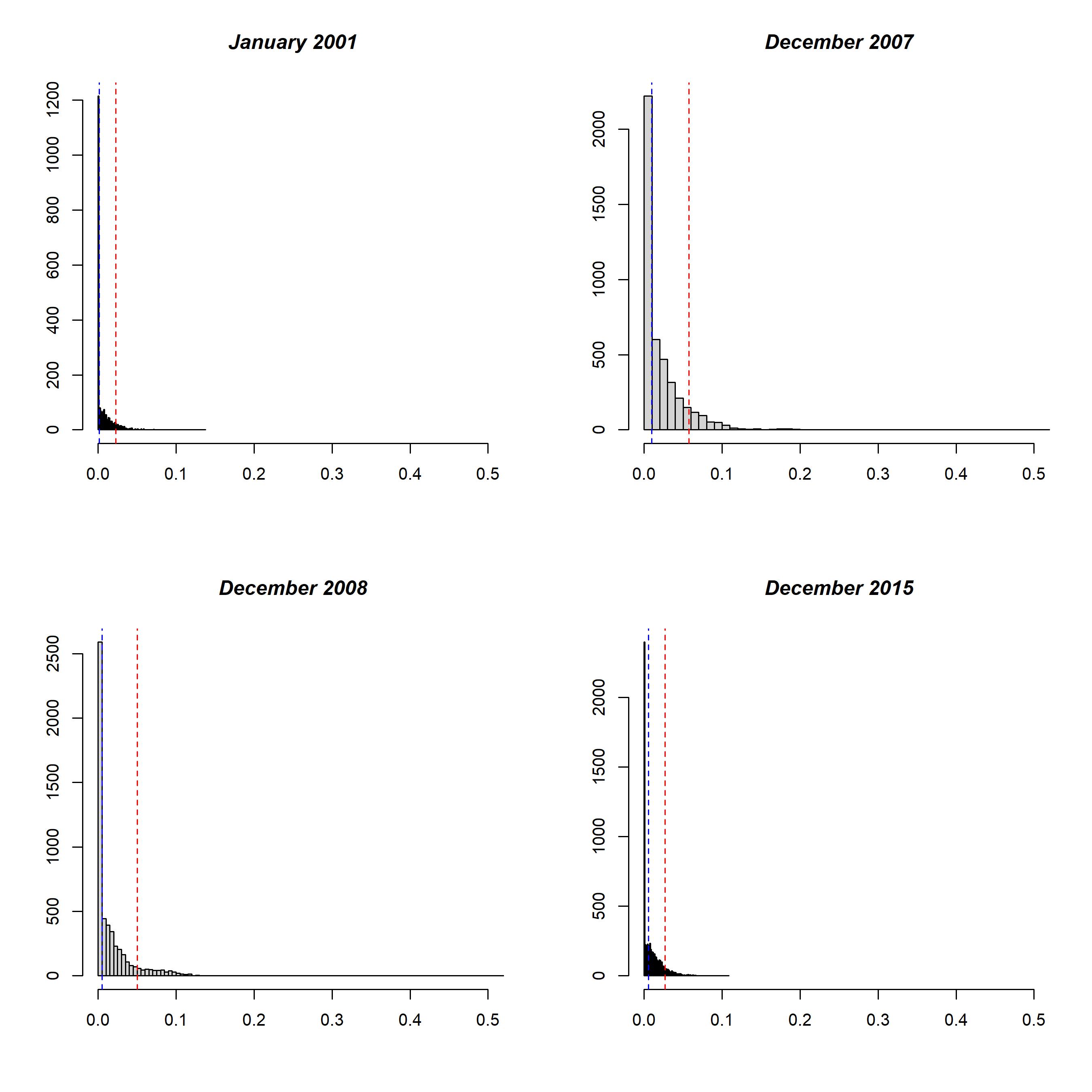}
	\caption{Distributions of edge weights at different time periods. In each figure, the dotted blue line represents the median, while the dotted red line is the 90\% quantile}
	\label{fig:Distr2}
\end{figure}

\begin{figure}[!h] 
	\centering
	\includegraphics[width=10cm,height=10cm]{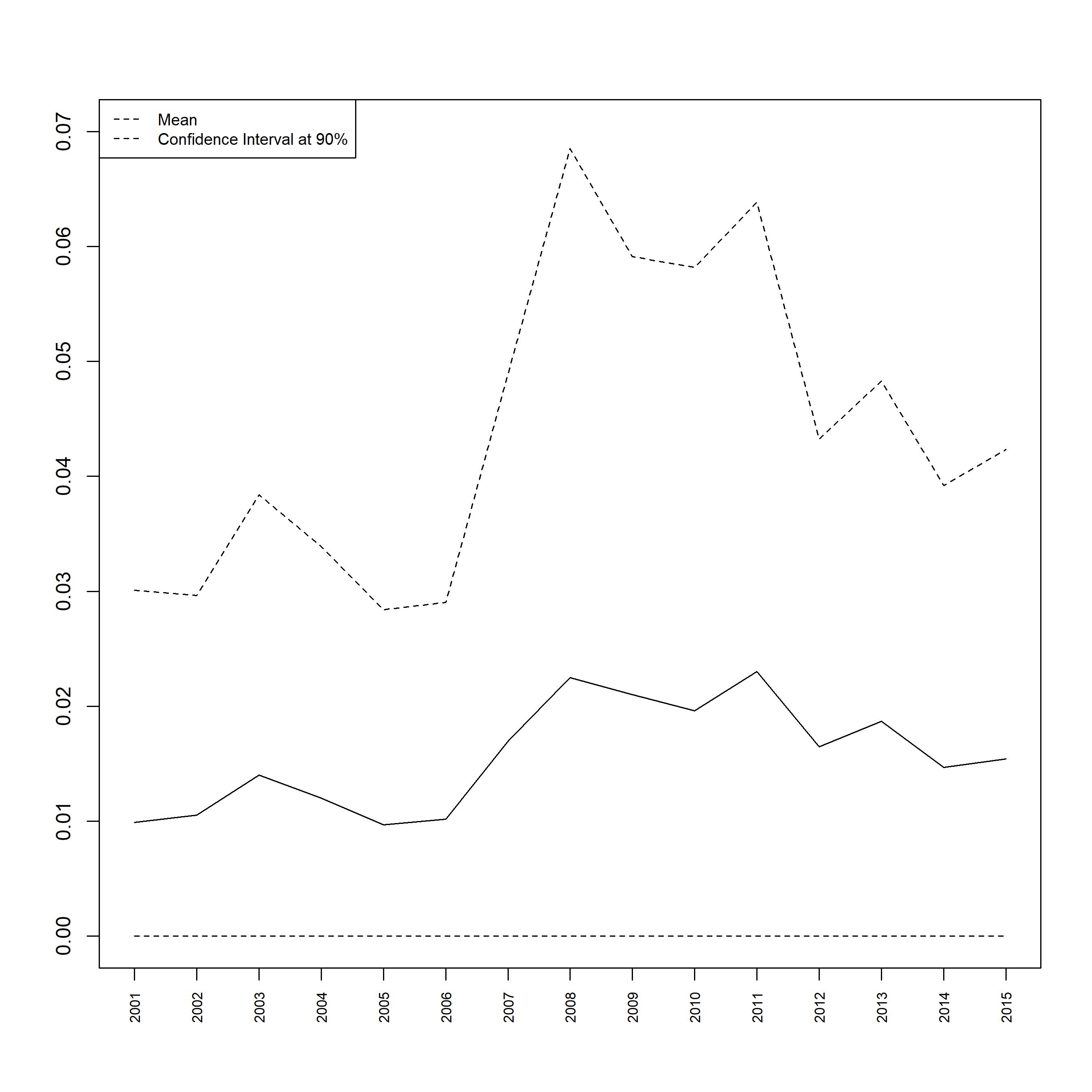}
	\caption{Mean and confidence interval at 90\% of weights distribution for each year.}
	\label{fig:Distr3}
\end{figure}

In Figure \ref{fig:Distr2}, we report the distributions of edges weights at four different time periods. The distributions capture the effects of the increased distress that already affected the market in 2007, i.e., both MES and ES slowly increased with respect to previous years. The average ratio between MES and ES increased in 2007 and reached a peak in the second half of 2008. \\
To give a visualization of weights' pattern on the whole period, we report in Figure \ref{fig:Distr3} mean and confidence interval at 90\% of weights distribution for each year. Another important effect can be detected in 2010 and 2011. We have an average increase in the weights because MES is growing faster than ES and, in particular, the differences in the patterns among the firms in the sample induce higher volatility. At the end of 2011 the standard deviation of the weights is higher than in previous periods. This is mainly because the sovereign debt crisis has affected five countries of the Eurozone in a more substantial way. The pattern of the weights also captures the reduction of contagion for other Eurozone countries, which diminished in the second half of 2012 because of the successful consolidation and implementation of structural reforms in the countries that were most at risk. \\

\subsection{Network Robustness Indicators and G-SII classification}

\begin{figure}[!h] 
	\centering
	\includegraphics[width=12cm,height=10cm]{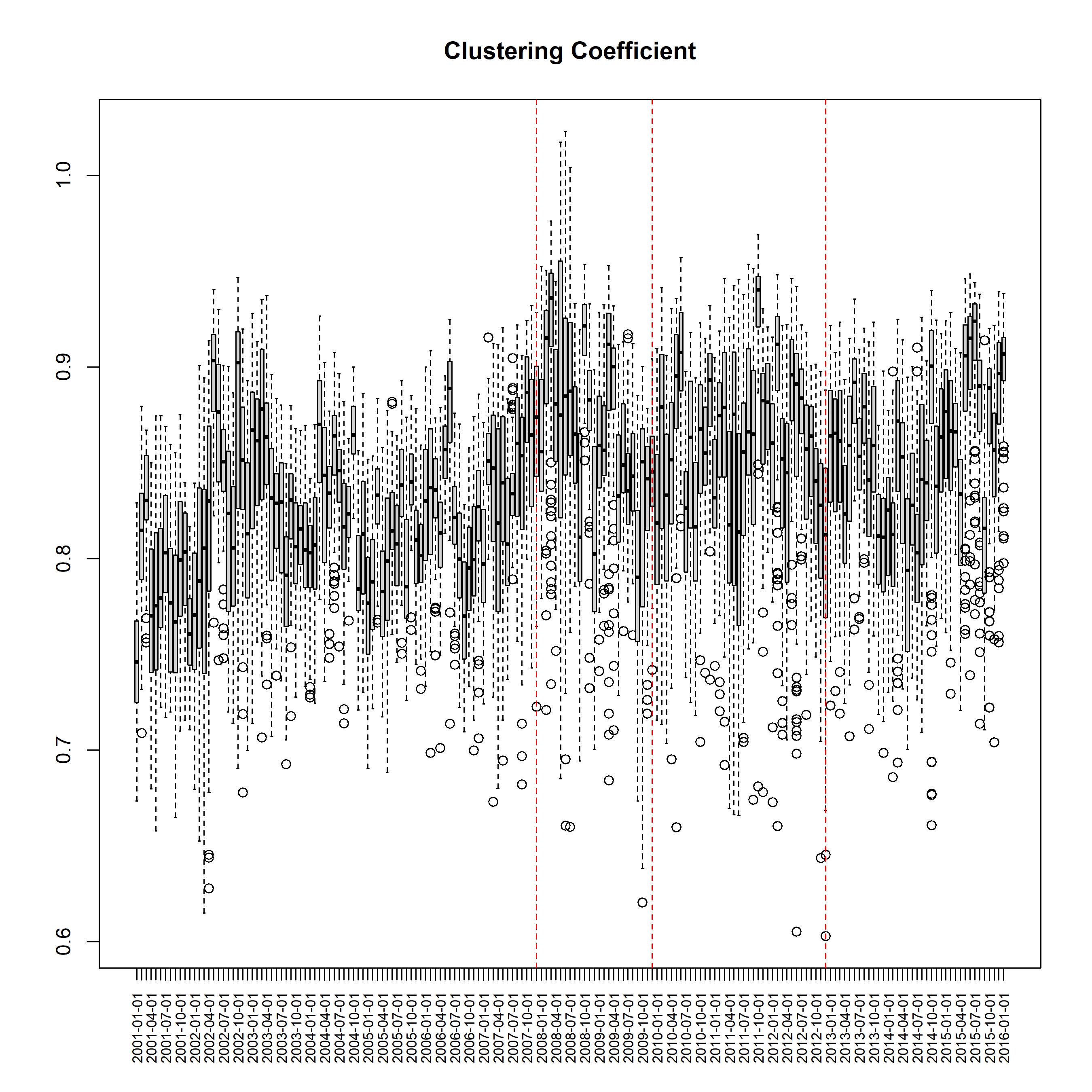}
	\caption{Distributions of clustering coefficients at different time periods. Four macro-periods that are characterized by different events are identified by the red dotted lines.}
	\label{fig:CDistr}
\end{figure}

To measure the level of interconnectedness in the network, we computed at each time period the clustering coefficient, a well known network indicator, introduced by \cite{Barrat_2004} for weighted networks. This indicator has been widely used in the literature as a measure of the state of stress of the financial market (see, e.g., \cite{Minoiu} and \cite{Tabak}). Since the considered time span covers very turbulent periods which affected countries in different ways and with different timing, we divided the time period in four different sub-periods. Following \cite{LoDuca2017} and \cite{Clemente2020a}, without any pretense of being exhaustive, we define four sub-periods, namely, \emph{Pre-crisis} (moving from January-2003  to December-2007), \emph{Lehman Brothers failure} (from January-2008 to December-2009), \emph{Sovereign debt crisis} (from  January-2010 to December-2012) and \emph{Post-crisis} (from January-2013 to December-2015). Distributions of clustering coefficients are reported in Figure \ref{fig:CDistr}. We observe that the median coefficient evolves consistently with the underlying financial events. It tends to be lower in calm periods and rise before crises. Sharper spikes occur around highly popular events that caused severe stress in the global financial system: for instance, the 2008-2009 episode stands out as an unusually large perturbation to the network.  Then, a decline is observed until 2010, when a greater focus emerged on sovereign debt in the Eurozone. A smaller peak is observed during 2011 because of the presence in the sample of several insurance companies from countries that were most severely affected by the sovereign debt crisis. \\ 

\begin{figure}[!h] 
	\centering
	\includegraphics[width=12cm,height=10cm]{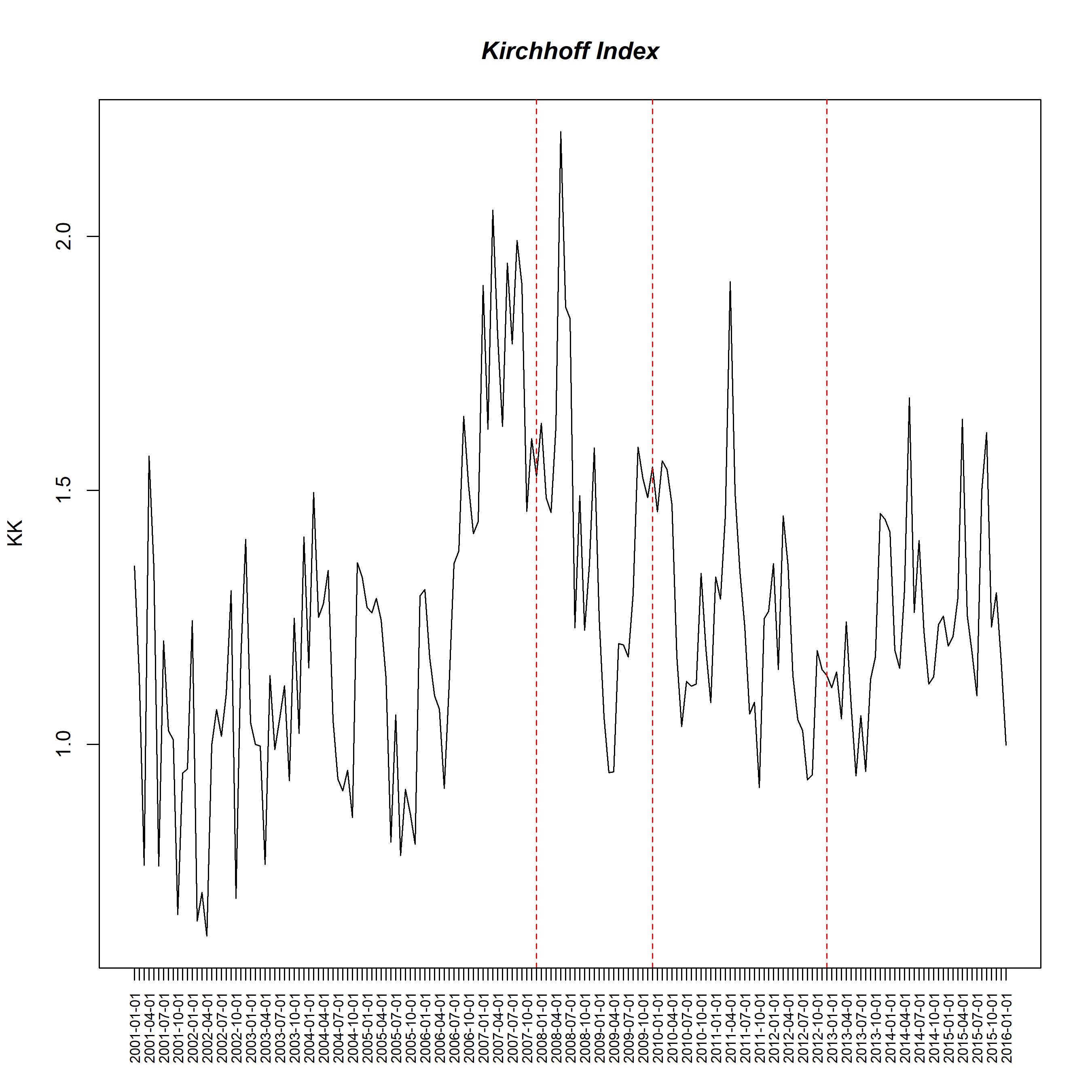}
	\caption{Pattern of normalized weighted Kirchhoff Index $K^{W}_{N}(G_{t})$  at different time periods $t$. Four macro-periods that are characterized by different events are identified by the red dotted lines.}
	\label{fig:Kir}
\end{figure}

To assess the relevance of each firm on the system, we compute the Weighted Effective Resistance Centrality for each vertex. Indeed, the aim is to catch the importance of each firm by measuring the relative variation of the normalized weighted Kirchhoff Index given by the removal of that vertex from the network. We repeat the procedure for each time period and we rank firms according to this index. Since the edges' weights are based on market-based measures, which are forward-looking by nature, our purpose is to identify which are the most relevant firms in a systemic risk perspective. Before than focusing on the behavior of each firm, we provide in Figure \ref{fig:Kir} the pattern of normalized weighted Kirchhoff Index over time computed at network level. Periods (as 2007-2008 and 2011) with a higher levels of interconnection in Figure \ref{fig:CDistr} are characterized by a network with a lower degree of robustness in Figure \ref{fig:Kir}. We remind indeed that a higher value of $K^{W}_{N}(G_{t})$ means a higher vulnerability of the network. In other words, in these periods, the failure of a firm can have a larger impact on the whole market.

Now, we focus on the behavior of each insurance company in terms of Weighted Effective Resistance Centrality (see Table \ref{SIFIClust}). In particular, we rank firms in decreasing order according to the average assumed by this index for each of the four sub-periods previously defined. Since we are using a measure of robustness computed here on a network that implicitly takes into account the tail dependence between equity' returns, we can interpret firms in the top quantile as the most interconnected  in the network. We compare them with the insurance companies that have been classified as G-SIIs (reported in italics in the Table). Since Aegon has replaced Generali Assicurazioni on the list in November 2015, we label as G-SIIs all firms that belong to the list at any time during the period. \\

\begin{table}[htbp]
	{\footnotesize \resizebox{1.05\textwidth}{!}{
			\centering
			\begin{tabular}{|c|c|c|c| c||}
				\hline\hline
				\multicolumn{5}{c||}{Top insurers according to higher average of the weighted effective resistance centrality} \\ \hline\hline
				\emph{Pre-crisis} & \emph{Lehman failure} & \emph{Sovereign debt crisis} & \emph{Post-crisis}  & \emph{All periods} \\ \hline		\hline
				\emph{AXA}	& Sumitomo Life & Legal \& General & Berkshire Hathaway	& \emph{MetLife}  \\
				\emph{MetLife} &\emph{MetLife} &	American Financial Group &	RGA &	\emph{AXA} \\
				\emph{AIG}	& \emph{AIG}	&	Berkshire Hathaway	& \emph{AIG} &		Massachusetts Mutual \\
				\emph{Prudential Financial} & \emph{AXA} &	State Farm &	Allstate &	RGA \\
				Massachusetts Mutual &	Principal Financial &	Ageas &	State Farm &	Allstate \\
				\emph{Allianz} &\emph{Prudential Financial} &	\emph{Prudential Financial} &	Chubb & \emph{AIG} \\
				Allstate &	RGA &	Hannover Re & 	\emph{MetLife}&	\emph{Allianz} \\
				The Travellers Companies &	CNA	& Massachusetts Mutual	&	Massachusetts Mutual Life &	\emph{Prudential Financial} \\
				Munich Re &	\emph{Allianz} & RGA&Principal Financial &	State Farm \\
				Swiss Re &	Zürich& \emph{MetLife} 	& 	Principal Financial	& Chubb \\
				Chubb &	\emph{Aviva} &Allstate	& \emph{Aegon}	& Legal \& General\\
				CNA	& Massachusetts Mutual &	CNA&	\emph{Aviva}	& Principal Financial \\
				XL Group &	XL Group &		RSA&	Ageas & \emph{Prudential Plc} \\
				Principal Financial	& State Farm & \emph{Prudential Plc}  &Legal \& General & CNA\\
				American Financial Group &	Sampo Oyj	&	Baloise	& \emph{AXA} &	Berkshire Hathaway\\
				\emph{Prudential Plc} &	Munich Re &	Principal Financial &Hannover Re &	Swiss Re \\
				State Farm & American Financial Group & Munich Re & 	Swiss Re&  Munich Re\\
				RGA &	Allstate &\emph{Allianz} &	Baloise	& XL Group \\
				Baloise	& Chubb	&  Chubb &\emph{Allianz} &	Baloise \\
				RSA & The Travelers Companies	& Swiss Re &	\emph{Prudential Plc} &	Zürich \\
				Ageas &	CNP Assurances & \emph{AXA} & \emph{Assicurazioni Generali} &	\emph{Aviva} \\
				Legal \& General&	Old Mutual	&\emph{AIG}	& CNA &	Hannover Re \\
				Zürich & \emph{Prudential Plc}	& Scor &	Zürich & Ageas\\
				\emph{Assicurazioni Generali} &	Swiss Re &	\emph{Argon} &	Groupama &	\emph{Assicurazioni Generali} \\
				Aviva &	Helvetia &	Zürich &	Sampo Oyj	& \emph{Aegon} \\
				Hannover Re	& \emph{Assicurazioni Generali} & \emph{Aviva} & Helvetia &	Sumitomo Life\\\hline \hline
				\multicolumn{5}{c||}{Ranking of G-SII that do not belong to the top quartile}\\ \hline\hline
				\emph{Pre-crisis} & \emph{Lehman failure} & \emph{Sovereign debt crisis} & \emph{Post-crisis} & \emph{All periods} \\ \hline	\\ \hline			
                Aegon (38) & Aegon (85) & Assicurazioni Generali (57) & Ping An (96) & Ping An (96) \\
                Ping An (73) & Ping An (95) & Ping An (101) & & \\ \hline\hline

			\end{tabular}%
	}}
	\caption{Top insurers (first quartile) in terms of average weighted effective resistance centrality. The insurers that are classified as G-SIIs are shown in italics. In the bottom part of the table we report the rankings of the G-SIIs that do not belong to the first quartile of the distribution.} 
	\label{SIFIClust}%
\end{table}%

It is worth pointing out that the methodology used by IAIS for identifying systemic important insurers is based on various characteristics, but the sizes of the firms and their interconnectedness with the system are two relevant issues. However, other points, as global activity, asset liquidation and substitutability, have been considered (see, e.g., \cite{EIOPA2017} and \cite{IAIS2013}). Hence, in our comparison, we are taking into account only a single aspect of this classification, namely, we deal with interconnections measured in the tails and we do not intend to propose a new methodology for assessing a G-SII.
Despite this clarification, it is interesting to note that insurance companies, that are classified as G-SIIs, also have a relevant impact on the robustness of the network. For instance, considering the whole period, we notice that Axa, MetLife, AIG, Allianz and Prudential belong to the top ten of our ranking. Additionally we also observe that over the last periods Aegon increased its ranking reaching higher levels of importance in the network than Assicurazioni Generali. This behavior seems in line with the IAIS classification that replaced Assicurazioni Generali with Aegon in the list of G-SII. \\
Main exception is represented by Ping An Insurance, that, although is classified as a G-SII, belongs to very low quartiles of the distribution of the Effective Weighted Resistance Centrality in all period. However, the presence of Ping An Insurance in the G-SII list is mainly justified by its size\footnote{The company has been classified as the third insurer in the world in terms of net premiums and in the top ten in terms of non-banking assets by AM Best in 2018.} that is not considered in our approach. \\
It is also noticeable that these results are in line with \cite{Denko}. Although the authors in \cite{Denko} used a smaller sample of firms and a different approach based on copula Garch models and minimum spanning tree, common points between the two papers are represented by the use of market-based indicators computed on equity returns and by the application of methodologies to catch the dependence in the tails. As in our case, the authors confirmed the relevance of several firms that belong to the list (as Axa, Allianz, Aegon, Aviva and Prudential) and the presence of other firms (as Zürich and Legal \& General) not included in the list. \\
We also notice that in our analysis some relevant firms of the U.S. market (see, e.g., Berkshire Hathaway and State Farm) appear as top players in terms of robustness of the network. \\
Finally, a long discussion regarded the systemic relevance of reinsurance companies as well as the possible inclusion of these firms in the G-SII list (see, e.g., \cite{ESRB}, Annex 4). Although we are not considering here risk transferring between insurers and reinsurers, it is interesting to note the presence of the largest reinsurance companies (Munich Re, Swiss Re and Hannover Re) in Table \ref{SIFIClust}. We can interpret this presence as a confirmation of the relevance of these firms in the market.

\section{Conclusions}
\label{sec:conc}
Systemic risk is an issue widely studied in the literature. In this context, a relevant topic that has been addressed is the need to tighten up supervision on global systemically important financial institutions. We focus in this paper on the insurance sector and we provide a methodology based on a combination of market-based measures and a network approach for assessing relevant firms in the market. \\
In particular, the aim is to propose a specific network indicator, the Weighted Effective Resistance Centrality, aimed at catching which is the effect on the robustness of the network of the removal of a specific firm. \\
The empirical analysis shows how the insurance companies, classified as G-SIIs by the IAIS, are identified as relevant firms by our indicator with very few exceptions. It is worth pointing out how the proposal is not an alternative with respect to the methodology provided by IAIS, but, instead, can be considered a complement to the more traditional approaches that are based on balance sheet and regulatory data.

\section*{Acknowledgements}
This work has been sponsored by the Casualty Actuarial Society (CAS). 

\bibliographystyle{spmpsci}      
\bibliography{PaperCC}   
\bibliographystyle{amsplain}

\newpage
\appendix
\begin{table}
			\centering
			\begin{tabular}{|l|l|}
				
				\hline\hline
				Achmea & Aegon N.V. \\
				 Aflac Inc & Ageas SA/NV \\
				AIA Group Ltd & AIG Inc \\
				 AIG Europe Ltd & Alleghany Corp. \\
				Allianz SE & Allstate Corp \\
				American Equity Investment Life Insurance Co & American Financial Group Inc/OH \\
				American National Group, Inc. & Ms Amlin Ltd \\
				AmTrust Financial Services Inc & 
				Asahi Mutual Life Insurance Co  \\
				Assicurazioni Generali Spa & Assurant Inc \\ 
				Aviva & AXA \\
				Baloise & Berkshire Hathaway \\ 
				Brighthouse Financial Inc & 
				Cathay Financial Holding Co Ltd \\
				Chesnara PLC & 			China Life Insurance Company Ltd \\ 
				China Pacific Insurance & China Taiping Insurance Group Ltd \\
				Chubb & CNA Financial Corporation \\ 
				CNO Financial Group Inc & CNP Assurances \\
				Dai-ichi Life Holdings Inc &  Delta Lloyd NV \\ 
				Direct Line Insurance Group PLC & DB Insurance Company \\
				Fidelity and Guaranty Life Insurance Company & Fubon Insurance Co Ltd \\ 
				Genworth Financial & Gjensidige Forsikring            \\
				Groupama Holding SA & Guardian Life Insurance Inc \\
				Hannover Re &  Hanwha Life Insurance\\
				Hartford Life and Accident Insurance Co & HDFC Life Insurance Co Ltd \\ Helvetia Holding AG & Insurance Australia Group Ltd \\
				Japan Post Insurance Co Ltd & KB Insurance Co Ltd \\ 
				Legal \& General Group PLC  & Lincoln National Corp\\ 
				Manulife Financial Corp & Mapfre Insurance Co \\ 
				Markel Corporation & Massachusetts Mutual Life Insurance Company \\ Mediolanum Assicurazioni & Meiji Yasuda Life Insurance Company \\ MetLife & MS \& AD Insurance Group \\
				Munich Re & Mutual of Omaha Insurance \\ 
				National Life Group &  New China Life Insurance Co\\
				New York Life Insurance Co & NFU Mutual \\ 
				NN Group N.V. & Northwestern Mutual Life Insurance Company \\ Nuernberger Beteiligungs AG & Ohio National Life Insurance Co \\ 
				Old Mutual plc & OneAmerica Financial Partners, Inc \\
				Penn Mutual Life Insurance Co & Phoenix Group Holdings PLC \\ 
				PICC Property & Casualty Co Ltd \\
				Ping An Insurance Group Co of China Ltd & Power Financial Corporation \\ 
				Principal Insurance Co & Prudential Financial In \\ 
				Prudential PLC & PZU SA Insurance \\ 
				QBE Insurance Group & RGA \\ 
				RSA Insurance Group & Sampo Oyj \\ 
				Samsung Fire \& Marine Insurance  & Samsung Life Insurance Co Ltd \\ 
				SCOR SE & Shin Kong Insurance Co \\ 
				Societa Cattolica di Assicurazioni SC & Sompo International Insurance\\
				Sony Life Insurance Co & Standard Life Aberdeen plc \\ 
				State Farm Mutual Automobile Insurance Company & Storebrand ASA \\
				Sumitomo Life Insurance Company & Sun Life Financial Inc \\ 
				Suncorp Group Ltd & Swiss Life Group \\
				Swiss Reinsurance Company Ltd & Symetra Financial Corp \\ 
				T\&D Life Group & Talanx AG\\
				The Progressive Corporation & The Travelers Companies, Inc. \\
				Thrivent Financial For Lutherans & TIIA \\ 
				Tokio Marine Holdings Inc & Topdanmark A/S \\
				 Tryg A/S & UnipolSai Assicurazioni Spa \\
				Uniqa Insurance Group & Unum Group \\
				 Vienna Insurance & Voya Financial Inc \\
				XL Group Ltd & Zenkyoren (National Mutual Insurance Federation of Agricultural Cooperatives ) \\
				 Zürich Insurance Group &\\
				
				\hline\hline
			\end{tabular}
		
	\caption{List of insurance companies in the sample}
	\label{ListBank}%
\end{table}

%
%

\newpage

\end{document}